\documentclass[aps, prd, twocolumn, superscriptaddress, showpacs]{revtex4}
\usepackage{graphicx}
\usepackage{amssymb}
\usepackage{amsmath}
\usepackage{booktabs}

\AtBeginDocument{
\heavyrulewidth=.08em
\lightrulewidth=.05em
\cmidrulewidth=.03em
\belowrulesep=.65ex
\belowbottomsep=0pt
\aboverulesep=.4ex
\abovetopsep=0pt
\cmidrulesep=\doublerulesep
\cmidrulekern=.5em
\defaultaddspace=.5em
\tabcolsep=3pt
}

\begin{document}
\title{Electron-positron pair production in frequency modulated laser fields}

\author{C. Gong}
\affiliation{State Key Laboratory for GeoMechanics and Deep Underground Engineering, China University of Mining and Technology, Beijing 100083, China}

\author{Z. L. Li \footnote{zlli@cumtb.edu.cn}}
\affiliation{State Key Laboratory for GeoMechanics and Deep Underground Engineering, China University of Mining and Technology, Beijing 100083, China}
\affiliation{School of Science, China University of Mining and Technology, Beijing 100083, China}

\author{B. S. Xie \footnote{bsxie@bnu.edu.cn}}
\affiliation{Key Laboratory of Beam Technology of the Ministry of Education, and College of Nuclear Science and Technology, Beijing Normal University, Beijing 100875, China}
\affiliation{Beijing Radiation Center, Beijing 100875, China}

\author{Y. J. Li \footnote{lyj@aphy.iphy.ac.cn}}
\affiliation{State Key Laboratory for GeoMechanics and Deep Underground Engineering, China University of Mining and Technology, Beijing 100083, China}

\date{\today}

\begin{abstract}
  The momentum spectrum and the number density of created electron-positron pairs in a frequency modulated laser field are investigated using quantum kinetic equation. It is found that the momentum spectrum presents obvious interference pattern. This is an imprint of the frequency modulated field on the momentum spectrum, because the momentum peaks correspond to the pair production process by absorbing different frequency component photons. Moreover, the interference effect can also be understood qualitatively by analyzing turning point structures. The study of the pair number density shows that the number density is very sensitive to modulation parameters and can be enhanced by over two orders of magnitude for certain modulation parameters, which may provide a new way to increase the number of created electron-positron pairs in future experiments.

\end{abstract}

\maketitle

\section{INTRODUCTION}

Since Dirac predicted the existence of positron~\cite{Dirac1928}, Sauter found that the vacuum could be broken into electron-positron pairs in a very strong static electric field~\cite{Sauter1931}, and then Schwinger pointed out that to produce observable pairs the field strength should reach the critical field strength $E_{\mathrm{cr}}\thickapprox 10^{18}\mathrm{V/m}$ by calculating the pair creation rate in a constant field with proper time method~\cite{Schwinger1951}. In order to figure out how to convert energy into mass in a vacuum, many methods have been proposed~\cite{Piazza2012,Xie2017}, such as world-line instanton~\cite{Dunne2005,Dunne2006}, computation quantum field theory~\cite{Lv2013,Jing2013,Su2012,Gong2018}, Wentzel-Kramers-Brillouin (WKB) approximation~\cite{Kim2002,Kim2007}, and quantum kinetic method  ~\cite{Alkofer2001,Akkermans2012,Hebenstreit2010,Li2014,Li2015,Li2017}.

However, the current laser intensity $\sim10^{22}\mathrm{W/cm^2}$ is still much smaller than the laser intensity corresponding to the critical field strength $\sim10^{29}\mathrm{W/cm^2}$. Therefore, it is impossible to produce observable electron-positron pairs experimentally only by the Schwinger tunneling mechanism. Moreover, another mechanism which is called multiphoton pair production can also break the vacuum into electron-positron pairs by absorbing several high-energy photons~\cite{Mocken2010,Akal2014}. Unfortunately, current laser facilities cannot provide high enough energetic photons for creating observable pairs from vacuum as well. To overcome this limitation, several catalytic mechanisms are put forward to create observable electron-position pairs in subcritical conditions. For example, the dynamically assisted Schwinger mechanism~\cite{Li2014,Schutzhold2008,Nuriman2012} can significantly enhance pair production by effectively combining the above two pair creation mechanisms. Another way to enhance the number of created pairs is the use of frequency chirped fields~\cite{Dumlu2010,Min2013,Abdukerim2017,Olugh2019}. It shows that the number density of created particles can be improved several orders of magnitude for suitable chirp parameters. However, it should be noted that in previous works the chirp parameter is not limited. Thus the pair creation can be greatly enhanced by absorbing photons with extremely high frequency component of the chirped field. Moreover, the momentum distribution of created particles is very sensitive to the chirp parameter, so one can only give a qualitative understanding of the momentum spectrum instead of a quantitative one.

In this paper, we study the details about electron-positron pair creation in a sinusoidal frequency modulated electric field by solving quantum Vlasov equation (QVE) numerically. This type of field can also be seen as a sinusoidal frequency chirped electric field. By giving a relatively reasonable restriction on modulation parameters, it is found that the number density of created pairs can still be enhanced by over two orders of magnitude for certain modulation parameters. Moreover, the interference pattern on the momentum spectrum is an imprint of the information carried by the modulated field and can be quantitatively understood by bridging momentum peaks and the frequency spectrum of the frequency modulated field. And a qualitative understanding of the interference effect is also given by analyzing turning point structures. In addition, the momentum and the number of created pairs can be artificially controlled by adjusting the modulation parameters of the frequency modulated electric field. By the way, this frequency modulated field may be also a good choice to check if Dirac vacuum can transmit the information carried by this field \cite{Su2019}.

This paper is organized as follows: In Sec.~\ref{sec:two} we briefly introduce the method of QVE which permits us to calculate the momentum distribution and the number density of created particles. In Sec.~\ref{sec:three}, we study the relation between the momentum peak and the laser frequency component of the frequency modulated field, and reanalyze the momentum spectrum within a semiclassical method. Moreover, we consider the number density of created particles for different modulation parameters. Section~\ref{sec:four} is a summary about this work.

\section{THE QUANTUM VLASOV EQUATION AND THE EXTERNAL FIELD}
\label{sec:two}

 The background laser field we used is a spatially homogeneous but time-dependent electric field $\mathbf{E}(t)=(0,0,E(t))$, and the corresponding vector potential is $\mathbf{A}(t)=(0,0,A(t))$ with $E(t)=-\dot{A}(t)$. Starting from the Dirac equation in the above field, one can derive the quantum Vlasov equation satisfied by the one-particle momentum distribution function $f(\mathbf{p},t)$ by a canonical time-dependent Bogoliubov transformation:
 \begin{eqnarray}\label{eq:qve}
 \frac{df(\mathbf{p},t)}{dt}=\frac{1}{2}\frac{eE(t)\varepsilon_{\perp}}{\Omega^{2}(\mathbf{p},t)} \int^{t}_{t_{0}}\!\!\!\!&&dt'\frac{eE(t')\varepsilon_{\perp}}{\Omega^{2}(\mathbf{p},t')}[1-2f(\mathbf{p},t')] \nonumber \\
 &&\!\times\cos\Big[2\int^{t}_{t'}d\tau \Omega(\mathbf{p},\tau)\Big],
 \end{eqnarray}
 where $-e$ and $m$ denotes the electron charge and mass, respectively. $\mathbf{p}=(\mathbf{p}_{\bot},p_{\parallel})$ is the canonical momentum, and $\varepsilon_{\bot}^{2}=m^{2}+\mathbf{p}_{\bot}^{2}$ represents the square of the perpendicular energy, $\Omega^{2}(\mathbf{p},t)=\varepsilon_{\bot}^{2}+k_{\parallel}^{2}(t)$ is the square of total energy, and $k_{\parallel}(t)=p_{\parallel}-eA(t)$ is the kinetic momentum along the direction of the electric field $E(t)$. In this paper, the natural unit $\hbar=c=1$ is used. Also note that the one-particle distribution function $f(\mathbf{p},t)$ is only a description of the generated real particles at $t\rightarrow+\infty$ where the external electric field is zero. Thus, we are just interested in the distribution function $f(\mathbf{p},\infty)$ and the particle number density $n(\infty)$.

 To solve Eq. (\ref{eq:qve}) numerically, one can introduce two auxiliary variables
 $u(\mathbf{p},t)=\int^{t}_{t_{0}}dt'W(\mathbf{p},t')[1-2f(\mathbf{p},t')]\cos[2\Theta(\mathbf{p},t',t)]$ and $v(\mathbf{p},t)=\int^{t}_{t_{0}}dt'W(\mathbf{p},t')[1-2f(\mathbf{p},t')]\sin[2\Theta(\mathbf{p},t',t)]$.
 Then Eq. (\ref{eq:qve}) can be equivalently transformed into the following first-order ordinary differential equations:
 \begin{eqnarray}\label{eq:system}
 \frac{df(\mathbf{p},t)}{dt}&=&\frac{1}{2}W(\mathbf{p},t)u(\mathbf{p},t),
 \nonumber \\
 \frac{du(\mathbf{p},t)}{dt}&=&W(\mathbf{p},t)[1-2f(\mathbf{p},t)]-2\Omega(\mathbf{p},t)v(\mathbf{p},t),\;
 \\
 \frac{dv(\mathbf{p},t)}{dt}&=&2\Omega(\mathbf{p},t)u(\mathbf{p},t).\nonumber
 \end{eqnarray}
 where $W(\mathbf{p},t)=eE(t)\varepsilon_{\perp}/\Omega^{2}(\mathbf{p},t)$ and $\Theta(\mathbf{p},t',t)=\int^{t}_{t'}d\tau \Omega(\mathbf{p},\tau)$.

 Finally, the single particle distribution function $f(\mathbf{p},t)$ can be obtained by solving Eqs. (\ref{eq:system}) with the initial conditions $f(\mathbf{p},-\infty)=u(\mathbf{p},-\infty)=v(\mathbf{p},-\infty)=0$. Then, the number density of created particles per $d^2p_\perp/(2\pi)^2$ evolving over time reads:
 \begin{equation}
 n(t)=2\int\frac{d p_\parallel}{2\pi}f(p_\parallel,t),
 \end{equation}
 where the factor $2$ comes from the degeneracy of electrons.

 In this paper, the configuration of the frequency modulated electric field is
 \begin{equation}\label{eq:filed}
 E(t)=E_{0} e^{-\frac{t^{2}}{2\tau^{2}}} \cos(\omega t+b\sin(\omega_{m}t)),
 \end{equation}
 where $E_{0}$, $\omega$ and $\tau$ are the strength, the frequency and the pulse duration of the electric field, respectively, $\omega_{m}$ is called modulated frequency and $b$ is a parameter which measures the degree of modulation.

\begin{figure}[!ht]
   \centering
   \includegraphics[width=8.5cm]{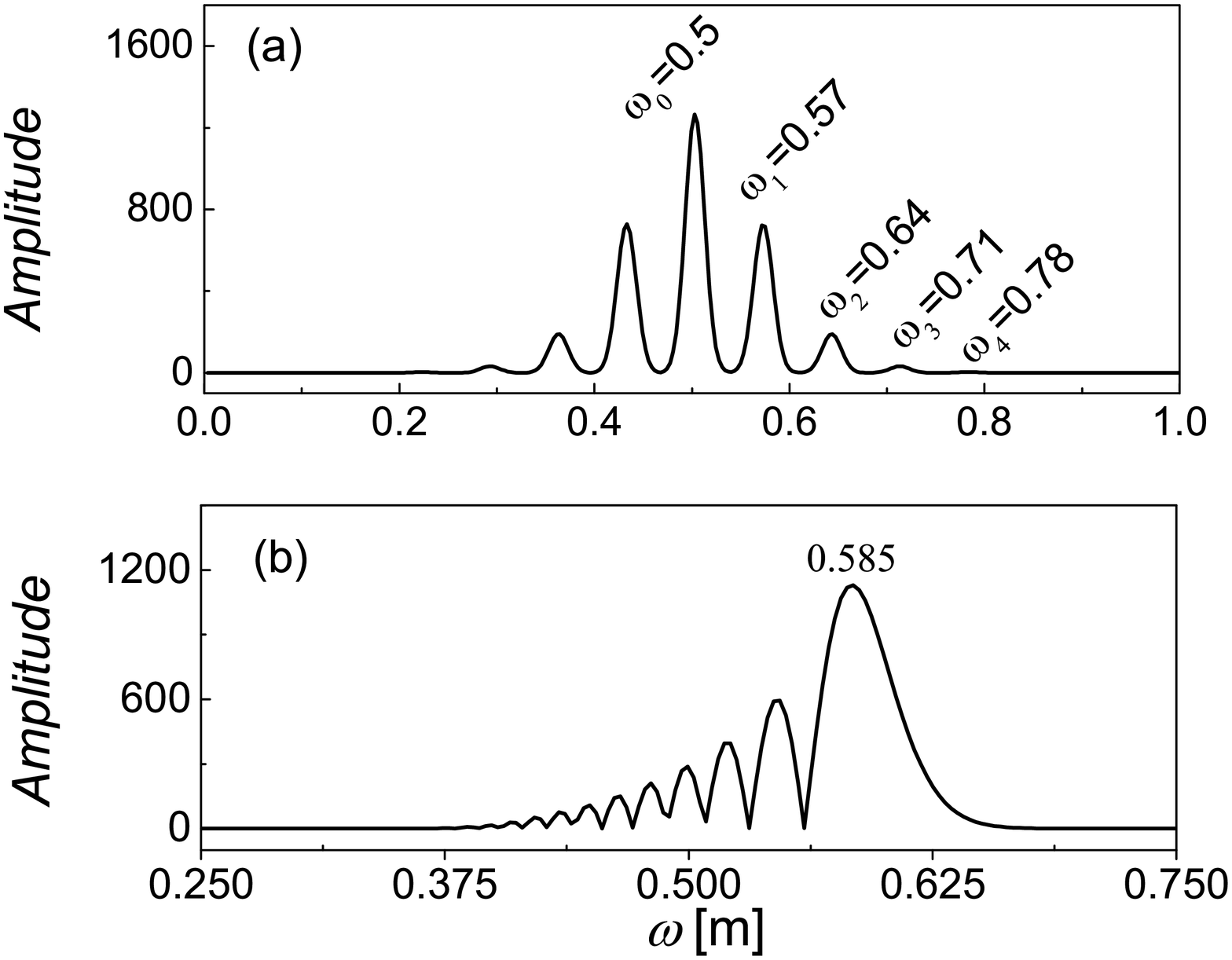}%
   \caption{The frequency spectra of the frequency modulated electric field with modulation parameters (a) $\omega_{m}=0.07m$, $b=1.0$ and (b) $\omega_{m}=0.01m$, $b=9.52$. Some typical values of frequency peaks are labeled on both panels. Other field parameters are $E_{0}=0.1E_{\mathrm{cr}}$, $\omega=0.5m$ and $\tau=100/m$.
   \label{fig:frequency}}
 \end{figure}

To help explain the following numerical results, we depict some typical frequency spectrum of the field (\ref{eq:filed}) in Fig. \ref{fig:frequency}. In Fig. \ref{fig:frequency} (a), the modulation parameters are $\omega_{m}=0.07m$ and $b=1$, and some frequency components are labeled as $\omega_{0}\sim\omega_{4}$. If we chose $\omega_{m}=0.1m$, then the frequency components which are larger than the original frequency are $\omega_{5}=0.6m$, $\omega_{6}=0.7m$, $\omega_{7}=0.8m$ and $\omega_{8}=0.9m$. For a large degree of modulation, the frequency spectrum is shown in Fig. \ref{fig:frequency} (b), where the modulation parameters are $\omega_{m}=0.01m$ and $b=9.52$. One can see that the original frequency are not the main frequency any more.

\section{NUMERICAL SIMULATIONS}
\label{sec:three}

\subsection{Momentum spectrum}

Generally, there are two different mechanisms that can induce electron-positron pair creation from vacuum. One is the Schwinger effect (tunneling mechanism) and the other one is the multiphoton absorption. These two mechanisms can be well distinguished by the adiabaticity parameter $\gamma=m\omega/|e|E_{0}$ \cite{Brezin1970}. For the former one $\gamma\ll1$ and for the latter one $\gamma\gg1$. While in this paper we are focused on the interesting intermediate regime $\gamma\sim\mathcal{O}(1)$. For instance, the adiabaticity parameter $\gamma=5$ for the modulated field with $E_{0}=0.1E_{\mathrm{cr}}$ and $\omega=0.5m$.

\begin{figure*}[!ht]
   \centering
   \includegraphics[width=0.95\textwidth]{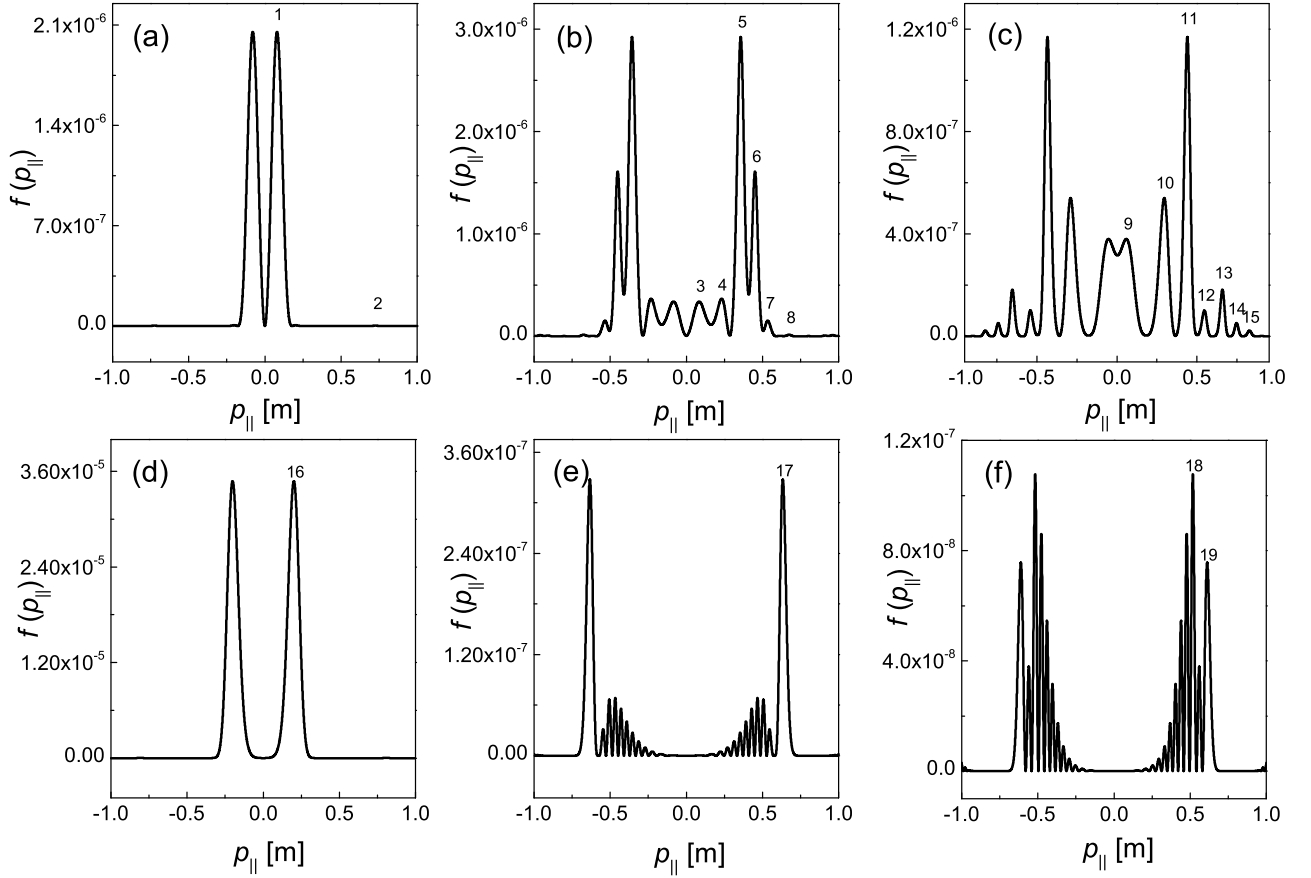}%
   \caption{The momentum spectra of created pairs in the modulated electric field with different modulation parameters. The modulation frequencies for (a), (b) and (c) are $\omega_{m}=0.00, 0.07m$ and $0.10m$, respectively. The degree of modulation is $b=1.0$. The modulation parameters ($\omega_{m}$, b) for (d), (e) and (f) are (0.01, 1.52), (0.01, 9.52) and (0.009, 9.52), respectively. Other field parameters are $E_{0}=0.1E_{\mathrm{cr}}$, $\omega=0.5m$ and $\tau=100/m$.
   \label{fig:momentum}}
 \end{figure*}

Figure \ref{fig:momentum} shows the momentum spectrum of created particles in frequency modulated fields with different modulation parameters. It can be seen that the momentum spectra in the modulated field have obvious interference pattern. To understand this interference effect, we show the peak positions of the momentum spectra labeled as $1$, $2$, ..., and $19$ on Fig. \ref{fig:momentum} in Tab. \ref{tab:pvalues}. According to the definition of total energy of a created electron, the total energy of created electron-positron pairs is
\begin{equation}\label{eq:enery}
 E(\mathbf{p})=2\sqrt{m_{\ast}^{2}+\mathbf{p}^{2}},
\end{equation}
where $m_{\ast}=m\sqrt{1+\frac{e^{2}}{m^{2}}\frac{E_{0}^{2}}{2\omega^{2}}}$ is called effective mass~\cite{Christian2014}. So the total energy $E_{1}$, $E_{2}$, ..., and $E_{19}$ which correspond to the momentum peak $p_{1}$, $p_{2}$, ..., and $p_{19}$ can be calculated by the above equation. For example, in Fig. \ref{fig:momentum} (a), we can obtain the total energy $E_{1}=2.02$ and $E_{2}=2.49$ according to Eq. (\ref{eq:enery}). Furthermore, we find that $E_{1}\approx4\omega_{0}$ and $E_{2}\approx5\omega_{0}$, so according to the energy conservation the created particles with momentum $p_{1}$ and $p_{2}$ correspond to four- and five-photon absorption process, respectively. And the number of created particles by four-photon absorption is much greater than that by five-photon absorption.

For Figs. \ref{fig:momentum} (b) and (c), the modulation frequency $\omega_{m}$ is $0.07m$ and $0.1m$, respectively. Due to the fact that $E_{3}, E_9\approx2.02\approx4\omega_{0}$ and $E_{8}, E_{14}\approx2.50=5\omega_{0}$, the created pairs with momentum $p_{3}$ and $p_{9}$ correspond to four-photon absorption and the created particles with momentum $p_{8}$ and $p_{14}$ correspond to five-photon absorption process. In addition to these momentum peaks corresponding to four- and five-photon absorption, one can also see that there are other momentum peaks in Figs. \ref{fig:momentum} (b) and (c). For Fig. \ref{fig:momentum} (b) where the modulation frequency $\omega_{m}=0.07m$, the total energy of created pairs $E_{4}\sim E_{7}$ corresponding to momentum peaks $p_{4}\sim p_{7}$ are $2.07$, $2.14$, $2.21$ and $2.28$, respectively. Based on energy conservation, these momentum peaks are still caused by four-photon absorption process, because $E_{4}\approx3\omega_{0}+\omega_{1}$, $E_{5}\approx3\omega_{0}+\omega_{2}$, $E_{6}\approx3\omega_{0}+\omega_{3}$ and $E_{7}\approx3\omega_{0}+\omega_{4}$. By the way, since the original frequency dominates the frequency components of the modulated electric field, a large number of photons with frequency $\omega_{0}$ is absorbed in pair creation. Similar to the above discussions, for Fig. \ref{fig:momentum} (c) with the modulation frequency $\omega_{m}=0.1m$, the total energy of created pairs $E_{10}\sim E_{13}$ corresponding to momentum peaks $p_{10}\sim p_{13}$ are $2.10$, $2.20$, $2.29$ and $2.41$, and $E_{10}\approx3\omega_{0}+\omega_{5}$, $E_{11}\approx3\omega_{0}+\omega_{6}$, $E_{12}\approx3\omega_{0}+\omega_{7}$, $E_{13}\approx3\omega_{0}+\omega_{8}$. So these momentum peaks also correspond to four-photon absorption process. Moreover, the total energy $E_{15}$ corresponding to the momentum peak $p_{15}$ is $2.61$. This peak is caused by a five-photon absorption process, because $E_{15}\approx4\omega_{0}+\omega_{6}$.

For the momentum spectrum of created pairs in a modulated field with a large degree of modulation $b$, see Figs. \ref{fig:momentum} (d), (e) and (f), the momentum peaks can also be determined by the frequency spectrum of the modulated field.  For instance, according to Eq. (\ref{eq:enery}), we have $E_{16}=2.05\approx4\times0.512$, $E_{17}=2.38\approx4\times 0.585$, $E_{18}=2.26\approx3\times 0.578+0.52$ and $E_{19}=2.35\approx4\times 0.578$, where $0.512$, $0.585$ and $0.578$ are exactly the dominant frequency components (see Fig. \ref{fig:point} (a), Fig. \ref{fig:frequency} (b) and Fig. \ref{fig:point} (b)), and $0.52$ is the frequency component corresponding to the third highest peak in Fig. \ref{fig:point} (b). Thus the momentum peaks $p_{16}$, $p_{17}$, $p_{18}$ and $p_{19}$ also correspond to four-photon absorption process. Moreover, we are surprised to find that the number of created pairs for $p_{18}$ is larger than that for $p_{19}$, though the momentum peak $p_{19}$ corresponds to the pair production by absorbing four photons with the dominant frequency component of the modulated electric field. This phenomenon can also be seen in Figs. \ref{fig:momentum} (b) and (c), where the highest momentum peak does not always correspond to the pair creation by absorbing four original frequency photons, though the original frequency dominates the frequency components. This finding will be explained in subsection \ref{C}.

Based on the analysis of these momentum peaks, we have a clear physical picture about how the interference effects came about. On one hand, we can use a frequency modulated laser beam to create electron-positron pairs with a predictable and particular momentum by suitable modulation parameters. On the other hand, the frequency components of a frequency modulated field can be well achieved by analyzing the momentum peaks, because the interference pattern is an direct imprint of the frequency modulated field on momentum spectra.

 \begin{table*}
   \begin{center}
   \caption{\label{tab:pvalues} The momentum values $p_{n}$ ($n$ from $0$ to $19$) of Fig. \ref{fig:momentum}. The unit is $m$.}
   \begin{tabular}{ccccccccccccccccccccc}
   \toprule
	  &\hspace{-0.3cm}$p_{1}$&$p_{2}$&$p_{3}$&$p_{4}$&$p_{5}$&$p_{6}$&$p_{7}$&$p_{8}$&$p_{9}$&$p_{10}$&
      $p_{11}$&$p_{12}$&$p_{13}$&$p_{14}$&$p_{15}$&$p_{16}$&$p_{17}$&$p_{18}$&$p_{19}$\!\!\!\!\!\!&\\
   \midrule
     &\hspace{-0.3cm}$0.080$&$0.728$&$0.083$&$0.232$&$0.356$&$0.451$&$0.535$&$0.731$&$0.059$&$0.294$&
      $0.440$&$0.547$&$0.658$&$0.746$&$0.828$&$0.196$&$0.633$&$0.517$&$0.611$\!\!\!\!\!\!&\\

   \bottomrule

   \end{tabular}
   \end{center}
\end{table*}

\subsection{Semiclassical analysis by turning-point structures}

The interference effects of momentum spectra and the number density of created electron-positron pairs can also be understood by phase-integral method, specifically, by analyzing the turning-point structure \cite{Cesim2010,Cesim2011}. The pair creation from vacuum in a spatially homogeneous and time-dependent modulated electric field is similar to the one-dimensional over-the-barrier scattering problem in quantum mechanics, and the momentum distribution function of created pairs can be obtained by the reflection coefficient.
\begin{equation}
 f(\mathbf{p})\!\approx\!\sum_{t_p}e^{-2K_\mathbf{p}^{p}}+\!\sum_{t_p\neq t_{p'}}\!2\cos(2\theta_\mathbf{p}^{(p,p')})(-1)^{p-p'}e^{-K_\mathbf{p}^{p}-K_\mathbf{p}^{p'}},
\end{equation}
with
\begin{eqnarray}
 K_\mathbf{p}^{p}&=&\left| \int_{t_p^*}^{t_p} \Omega(\mathbf{p},t)\, dt\, \right |, \quad
 \theta_\mathbf{p}^{(p,p')}=\left| \int_{\mathrm{Re}(t_p)}^{\mathrm{Re}(t_{p'})} \Omega(\mathbf{p},t)\, dt\, \right |, \nonumber
\end{eqnarray}
where $t_p$ and $t_{p'}$ represent different turning points which are the solutions of equation $\Omega(\mathbf{p},t)=0$. It is not hard to see that $K_\mathbf{p}^{p}$ and $\theta_\mathbf{p}^{(p,p')}$ determine the created pair's number and the degree of interference, respectively. To be specific, the particle number is dominated by the turning points which are nearest to the real time axis, while the degree of interference depends on the pair number of these turning points that are nearest to the real time axis. Therefore, the distribution of turning points on the complex time axis can be used to analyze the information of momentum spectra qualitatively, especially some specific momentum peaks.

 \begin{figure*}[!ht]
   \centering
   \includegraphics[width=0.3\textwidth]{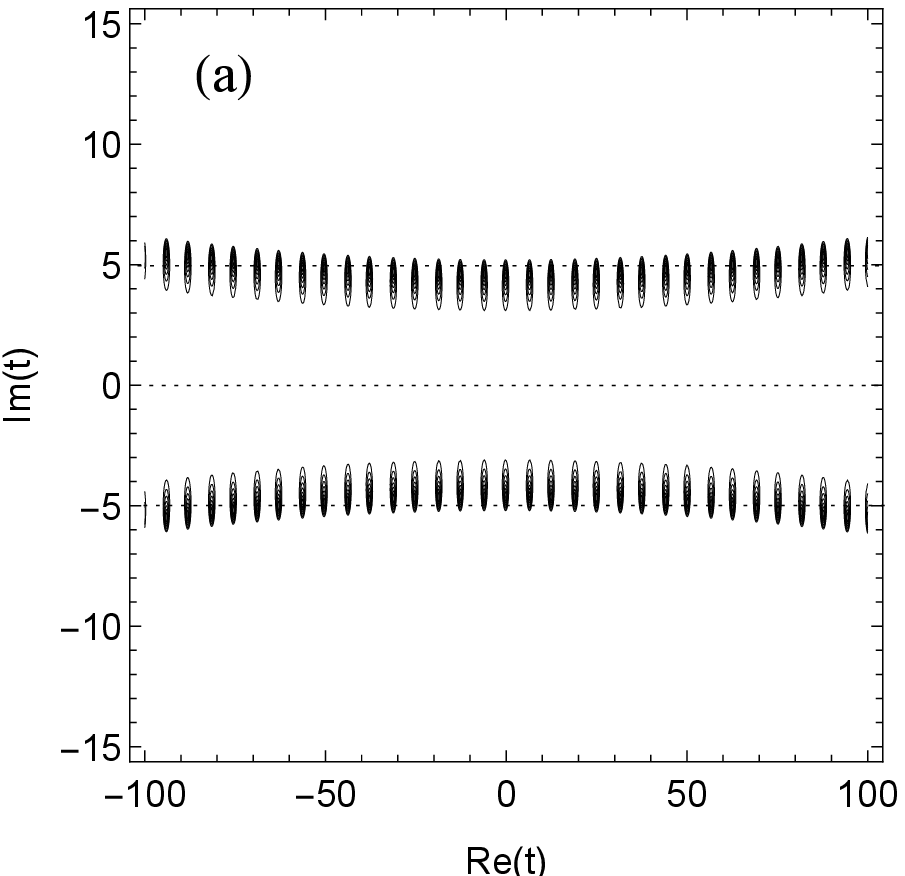}%
   \quad
   \quad
   \includegraphics[width=0.3\textwidth]{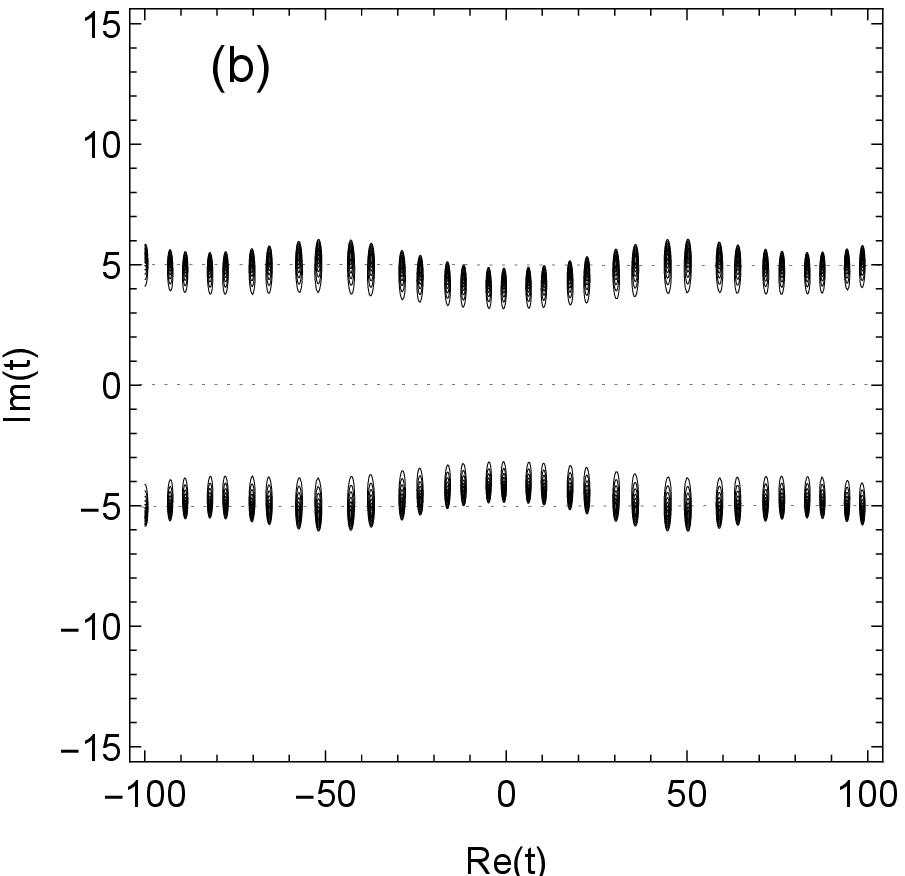}%
   \quad
   \quad
   \includegraphics[width=0.3\textwidth]{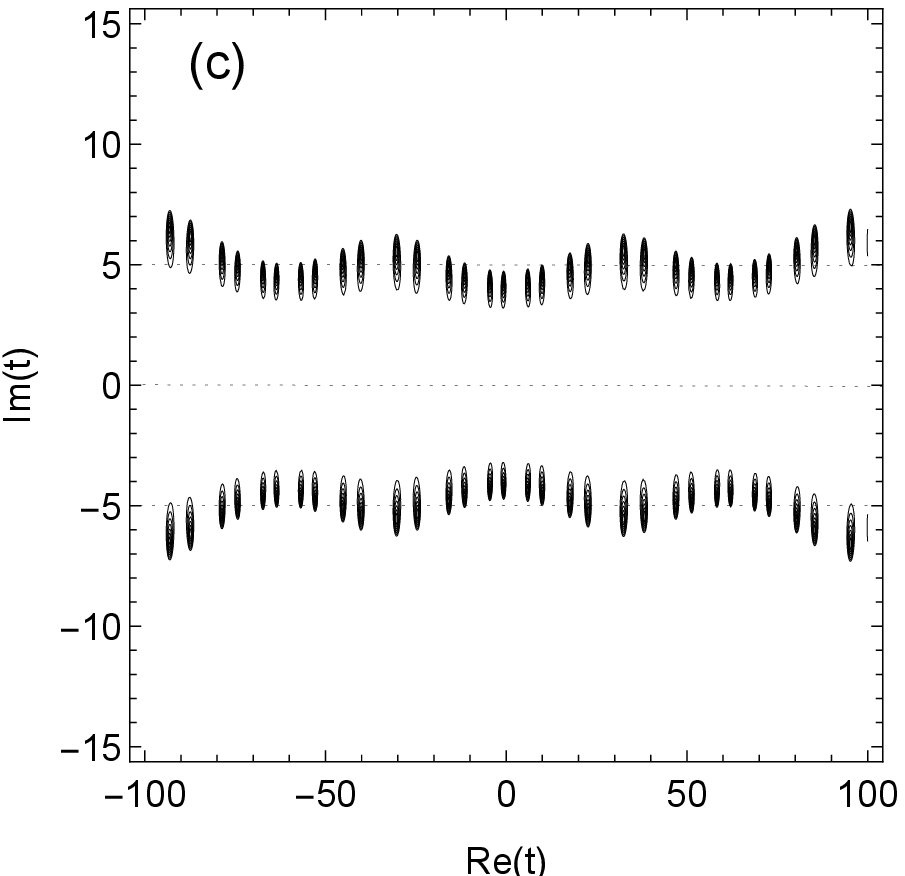}%

   \quad

   \includegraphics[width=0.3\textwidth]{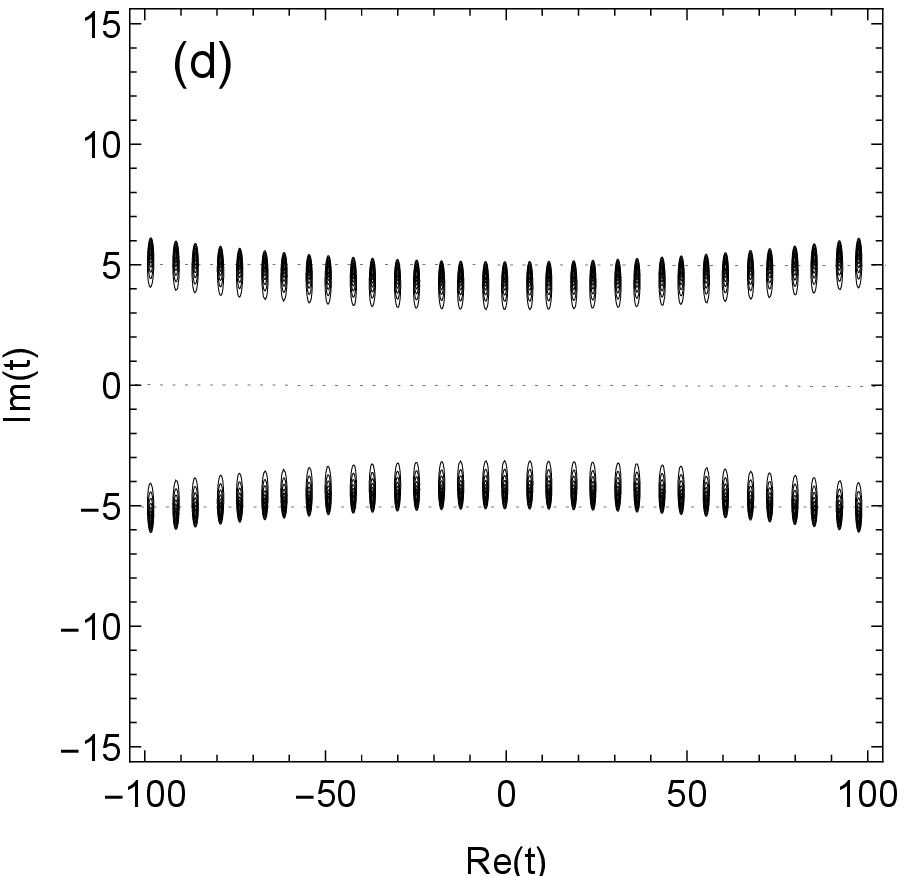}%
   \quad
   \quad
   \includegraphics[width=0.3\textwidth]{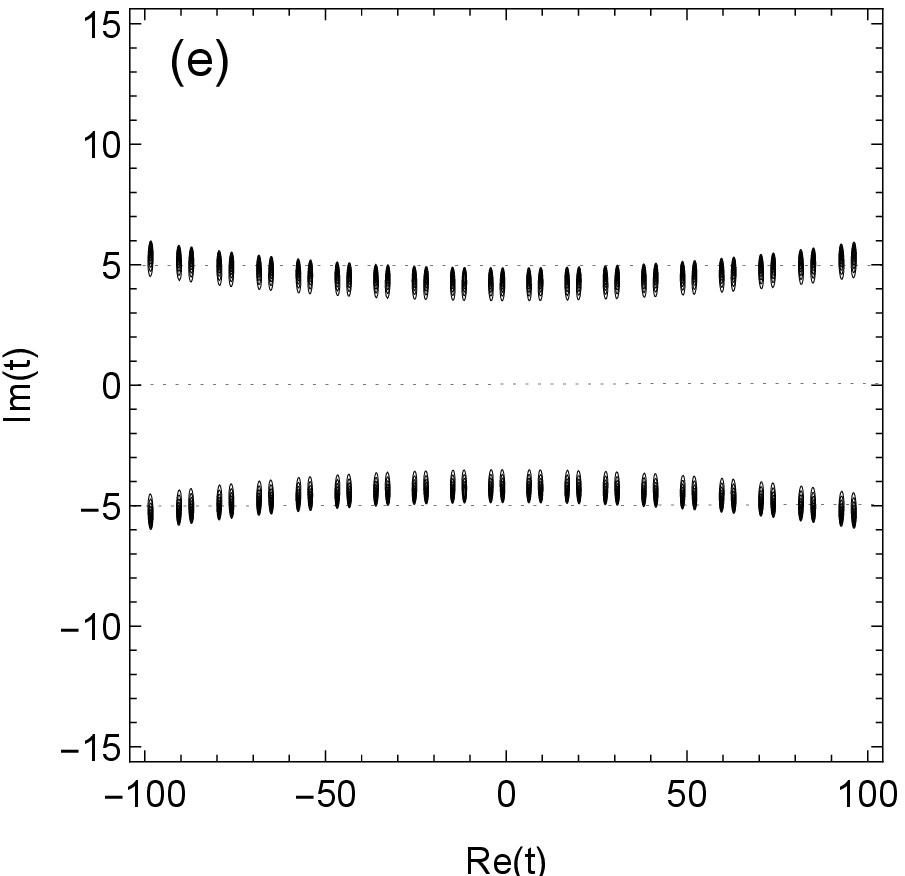}%
   \quad
   \quad
   \includegraphics[width=0.3\textwidth]{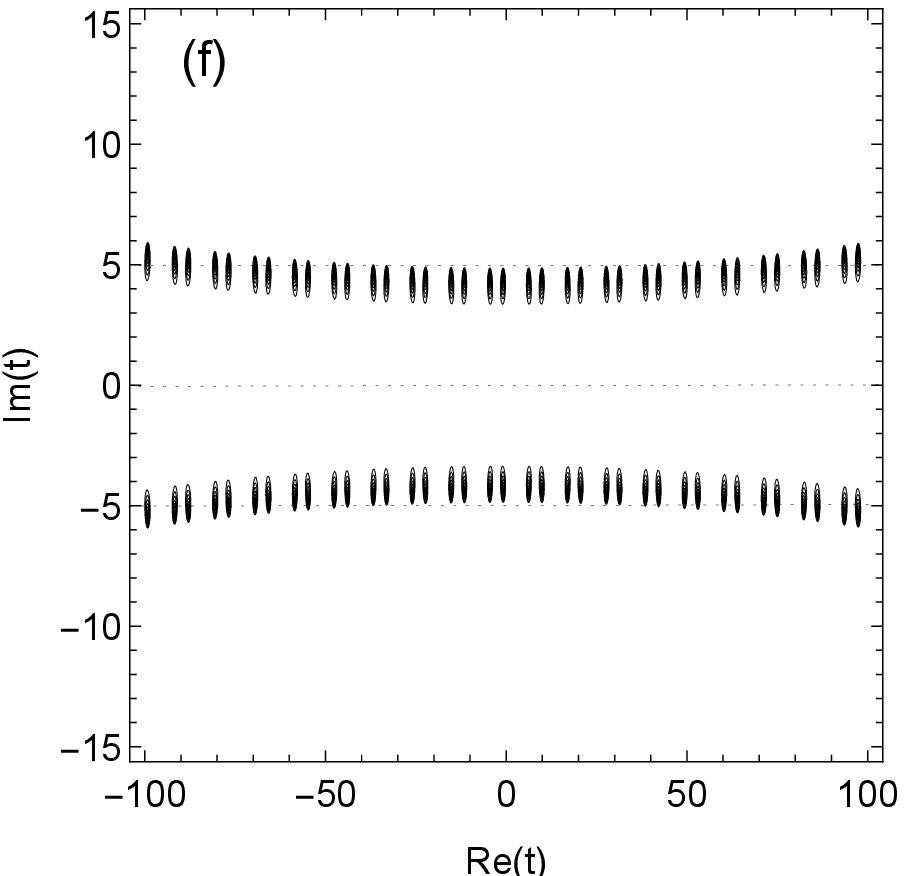}%
   \caption{Contour plots of $|\Omega (\mathbf{p},t)|^{2}$ in the complex $t$ plane, showing the location of turning points where $\Omega (\mathbf{p},t)=0$. For the upper three panels from left to right the modulation frequency and momentum correspond to $\omega_{m}=0$, $p_{1}=0.08m$; $\omega_{m}=0.07m$, $p_{5}=0.356m$; $\omega_{m}=0.1m$, $p_{11}=0.44m$, respectively, and $b=1.0$.  For the lower three panels from left to right the momentum values are $p_{16}$, $p_{17}$ and $p_{18}$, respectively. The modulation parameters are the same as in Figs. \ref{fig:point} (a), \ref{fig:frequency} (b), and \ref{fig:point} (b), respectively. Other field parameters are $E_{0}=0.1E_{\mathrm{cr}}$, $\omega=0.5m$ and $\tau=100/m$.
   \label{fig:turning}}
 \end{figure*}

When the electric field is modulated, the analytic expression for turning points can not be derived, so we give the distribution of these turning points by solving equation $\Omega(\mathbf{p},t)=0$ numerically. For the frequency modulated electric field with different modulation parameters, the turning points are depicted in Fig. \ref{fig:turning}. From Fig. \ref{fig:turning} (a) to Fig. \ref{fig:turning} (f), the distribution of turning points correspond to the momentum value $p_{1}$, $p_{5}$, $p_{11}$, $p_{16}$, $p_{17}$ and $p_{18}$, respectively. For the first row of Fig. \ref{fig:turning}, the modulation frequency $\omega_{m}$ is $0$, $0.07m$ and $0.1m$ from left to right, respectively. It can be seen that as the modulation frequency increases the periodicity of these turning points becomes shorter and shorter, and there will be more pair number of turning points close to the real time axis. Thus, the interference effect of the momentum spectrum becomes more and more obvious, see the first row of Fig. \ref{fig:momentum}. Furthermore, due to the fact that the turning points in Fig. \ref{fig:turning} (c) are closer to the real axis than that in Fig. \ref{fig:turning} (a), the magnitude of momentum peak $p_{11}$ in Fig. \ref{fig:momentum} (c) is smaller than that of $p_{1}$ in Fig. \ref{fig:momentum} (a). Of course, the magnitude of other momentum peaks can also be qualitatively analyzed by comparing the distance of turning points to the real time axis. For Fig. \ref{fig:turning} (d) $\sim$ (f) with the modulation frequency $\omega_m=0.01m$, $0.01m$ and $0.009m$, respectively, one can see that the period of turning points is not obvious, because the period of turning points depends on the modulation frequency and the modulation frequency in Fig. \ref{fig:turning} (d) $\sim$ (f) are much less than that in Figs. \ref{fig:turning} (b) and (c). Comparing the pair number of turning points nearest the real time axis between Fig. \ref{fig:turning} (e), (f) and Fig. \ref{fig:turning} (a), we can also find that the inference effect is more obvious for the former two figures than the latter one. Additionally, since the turning points in Fig. \ref{fig:turning} (e) are farther from the real axis than that in Fig. \ref{fig:turning} (d), the magnitude of momentum peaks in Fig. \ref{fig:momentum} (d) where $b=1.52$ is significantly larger than that in Fig. \ref{fig:momentum} (e) where $b=9.52$. And we will elaborate further on the specific reasons that affect the number of created particles in the next subsection.

\subsection{Pair number density}\label{C}

In this subsection, we study the number density of the created electron-positron pairs in frequency modulated electric fields. To keep the modulation within a reasonable range, we rewrite the modulated field (\ref{eq:filed}) as
\begin{equation}
E(t)=E_{0} e^{-\frac{t^{2}}{2\tau^{2}}} \cos(\omega_{\mathrm{eff}}t),
\end{equation}
where $\omega_{\mathrm{eff}}=\omega+\frac{b\sin(\omega_{m}t)}{t}$ is a time-dependent effective frequency, and set $|\frac{b\sin(\omega_{m}t)}{t}|\leq\alpha\omega$ with $0\leq\alpha<1$ for any time $t$. Because of $|\frac{b\sin(\omega_{m}t)}{t}|_{\max}=b\omega_{m}$ , we further obtain $b\omega_{m}\leq\alpha\omega$. Finally we have the constraint relation between the degree of modulation $b$ and the modulation frequency $\omega_{m}$ as
\begin{equation}\label{eq:divid}
 b\leq\frac{\alpha\omega}{\omega_{m}}.
\end{equation}

\begin{figure*}[!ht]
   \centering
   \includegraphics[width=0.5\textwidth]{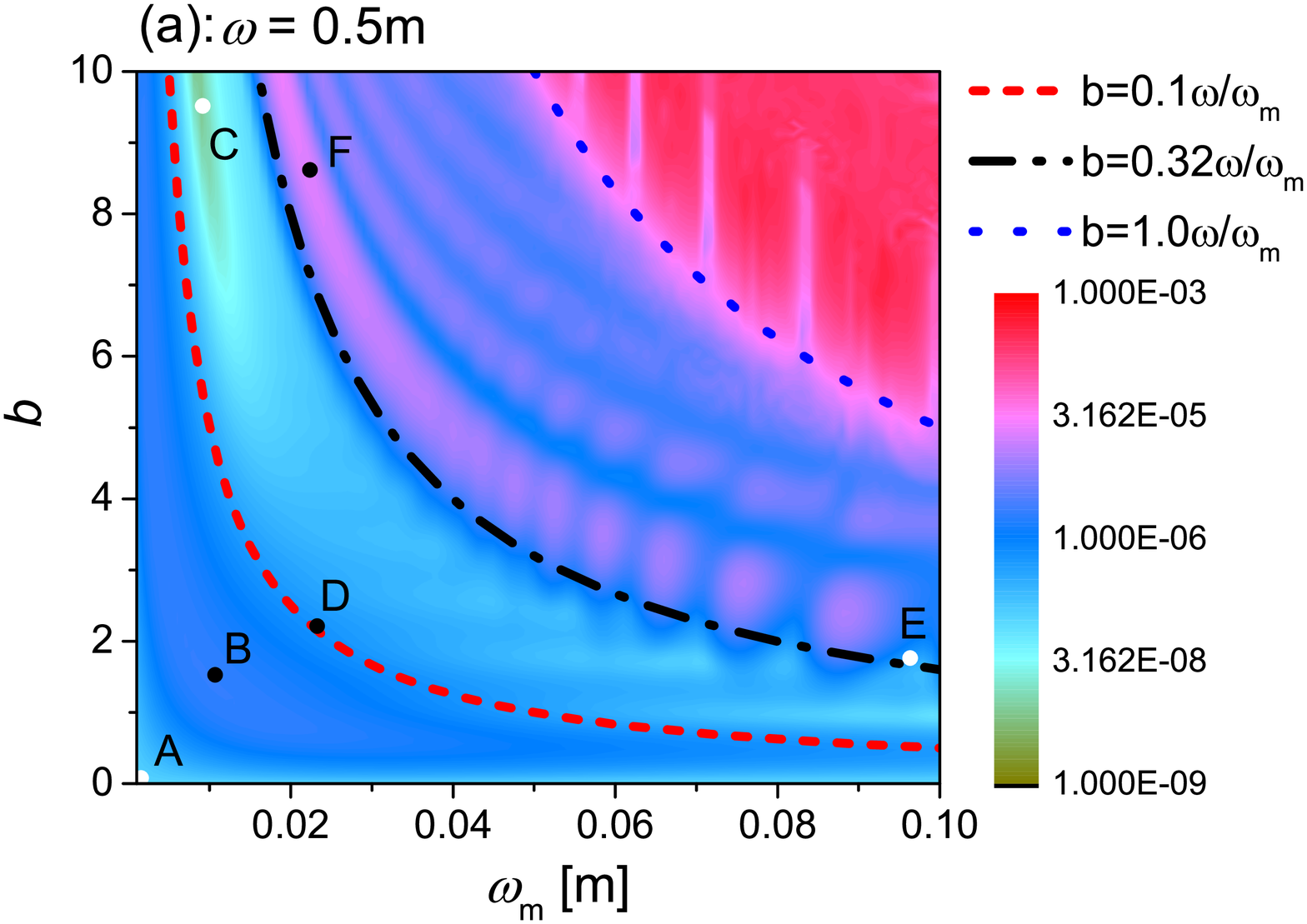}%
   \includegraphics[width=0.5\textwidth]{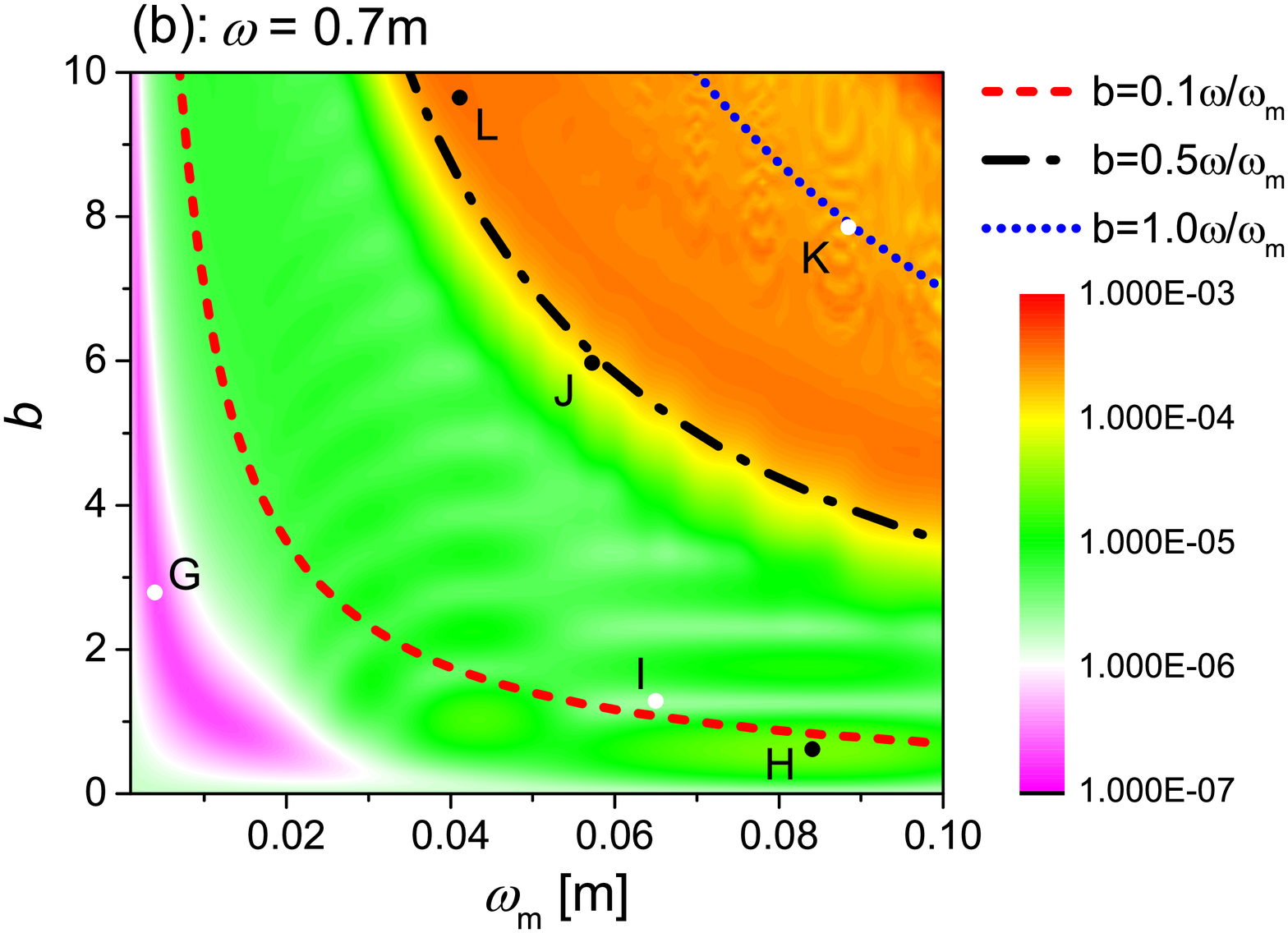}%
   \caption{Contour plot of the number density of created electron-positron pairs for different modulation parameters $(\omega_m, b)$. The red dashed, black dash-dot, and blue dotted line are the dividing line between the degree of frequency modulation. The white points $A$, $C$, $E$, $G$, $I$, $K$ represent the minimum values for each region, while the black points $B$, $D$, $F$, $H$, $J$, $L$ represent the maximum values. From left to right the values of the laser frequency are $\omega=0.5m$ and $\omega=0.7m$. Other field parameters are $E_{0}=0.1E_{\mathrm{cr}}$, $\tau=100/m$.
   \label{fig:number}}
\end{figure*}

\begin{table*}
   \begin{center}
   \caption{\label{tab:nmore} The number density for different selected sets of modulation parameters $(\omega_{m}, b)$. The laser frequency $\omega$ of these two sheets are $0.5m$ and $0.7m$ form top to bottom, respectively. The minimum and maximum values of the number density correspond to the points in Fig. \ref{fig:number}.}
   \begin{tabular}{ |c|c|cc|cc|cc| }
   \hline
	  $(\omega_{m}, b)$&$(0,0)$&$A(0,0)$&$B(0.01,1.52)$&$C(0.009,9.52)$&
      $D(0.023,2.24)$&$E(0.096,0.96)$&$F(0.022,8.64)$\\
   \hline
       $\mathrm{Number\;density}$& $1.04\times 10^{-7}$ & $1.04\times 10^{-7}$&$2.00\times 10^{-6}$& $7.63\times 10^{-9}$&$6.10\times 10^{-7}$ &
        $9.89\times 10^{-8}$&$2.03\times 10^{-5}$\\
   \hline\hline
      $(\omega_{m}, b)$&$(0,0)$&$G(0.004,2.8)$&$H(0.084,0.64)$&$I(0.065,1.28)$&
      $J(0.057,6.0)$&$K(0.088,7.9)$& $L(0.041,9.68)$\\
   \hline
      $\mathrm{Number\;density}$& $1.74\times 10^{-6}$&$2.02\times 10^{-7}$ & $2.93\times 10^{-5}$&$2.60\times 10^{-6}$ &
      $1.37\times 10^{-4}$&$9.89\times 10^{-5}$ & $3.50\times 10^{-4}$\\
   \hline
   \end{tabular}
   \end{center}
\end{table*}

To help study the effect of modulation parameters on the number density of created electron-positron pairs, we draw several typical curves (the red dashed, black dash-dot and blue dotted lines) according to Eq. (\ref{eq:divid}) on Fig. \ref{fig:number}, where shows the pair number density created in the frequency modulated electric field with different modulation parameters. Figures \ref{fig:number} (a) and (b) correspond to the laser frequency $\omega=0.5m$ and $0.7m$, respectively. And in both of them the degree of modulation $b$ varies from $0$ to $10$, and the modulation frequency $\omega_{m}$ varies from $0$ to $0.1m$ in order to ensure $\omega_{m}\ll\omega$. According to the red dashed line, i.e., $\alpha=0.1$, for any frequency modulated electric field with modulation parameters below the parameters of red dashed line, the maximum value of the effective frequency of the modulated electric field is no more than $\omega+0.1\omega$, which has the same order of magnitude as the laser frequency $\omega$ without modulation. Due to the same reason, for the black dash-dot line where $\alpha=0.32$ in Fig. \ref{fig:number} (a) and $\alpha=0.5$ in Fig. \ref{fig:number} (b), we can see that the number density of created particles divided by the black dash-dot line have an obvious change. It has the same change for the blue dotted line where $\alpha=1.0$.

From Fig. \ref{fig:number}, one can see that the number density of created pairs in several different areas which are separated by the typical lines has obvious minimum and maximum values. The minimum values are marked by white points $A$, $C$, $E$, $G$, $I$, $K$ and the maximum values are marked with black points $B$, $D$, $F$, $H$, $J$, $L$. The modulation parameters ($\omega_{m}, b$) and the number density of created pairs corresponding to each minimum and maximum values are shown in Tab. \ref{tab:nmore}. The result in the case of without frequency modulation, i.e., $\omega_{m}=b=0$, is also given.

\begin{figure}[!ht]
   \centering
   \includegraphics[width=8.5cm]{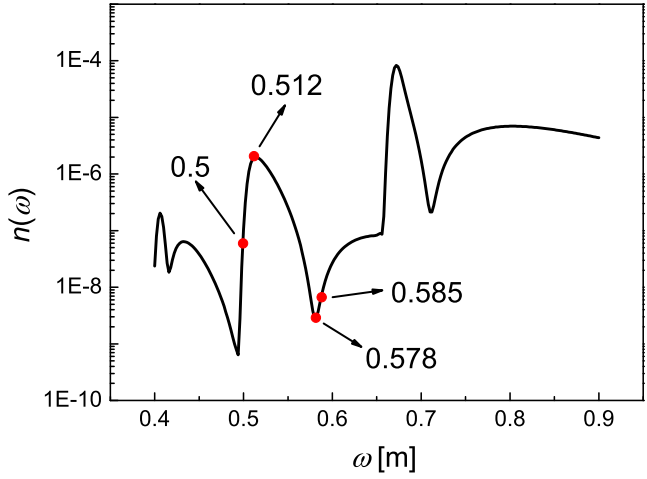}%
   \caption{The number density of created electron-positron pairs as a function of field frequency $\omega$. The oscillating structures are related to the $n$-photon thresholds. Other field parameters are $E_{0}=0.1E_{\mathrm{cr}}$ and $\tau=100/m$. Note that there is no frequency modulation, i.e., $b=0$.
   \label{fig:range}}
 \end{figure}

\begin{figure}[!ht]
   \centering
   \includegraphics[width=8.5cm]{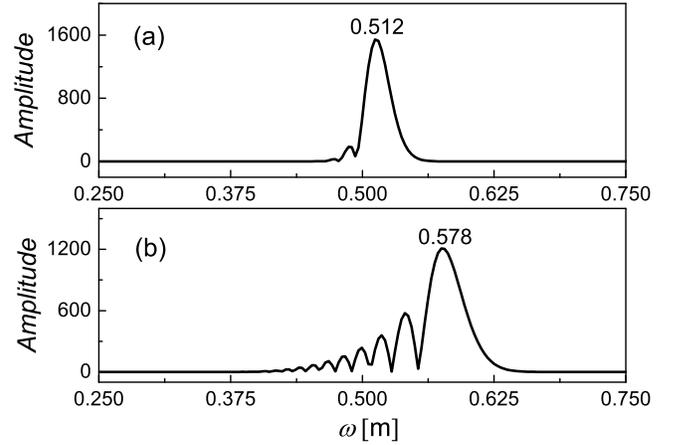}%
   \caption{The frequency spectra of the frequency modulated electric field corresponding to point $B$ and $C$ in Fig. \ref{fig:number} from top to bottom. The values of modulation parameter $(\omega_{m}, b)$ are $(0.01,1.52)$ for the upper panel and $(0.009,9.52)$ for the lower panel. And the values of dominant frequency peaks are shown. Other field parameters are $E_{0}=0.1E_{\mathrm{cr}}$, $\omega=0.5m$ and $\tau=100/m$.
   \label{fig:point}}
 \end{figure}

For Fig. \ref{fig:number} (a) where the laser frequency $\omega=0.5m$, when the laser frequency is not modulated, the number density of created particles is $1.04\times 10^{-7}$. While for $0\leq\alpha\leq0.1$, the range of modulation parameters $(\omega_{m}, b)$ is below the red dashed line, and the minimum and maximum values of the pair number density in this region are $A: 1.04\times 10^{-7}$ and $B: 2.00\times 10^{-6}$, respectively. It is found that the number density of created pairs corresponding to point $B$ is $20$ times greater than that without frequency modulation. For $0.1\leq\alpha\leq0.32$, the range of modulation parameters is between the black dash-dot line and the red dashed line, and the minimum and maximum values of the particle number density in this region are $C: 7.63\times 10^{-9}$ and $D: 6.1\times 10^{-7}$, respectively. Although the maximum value of the number density corresponding to point $D$ is larger than that without frequency modulation, it is smaller than that corresponding to point $B$ where the modulation parameters belong to $0\leq\alpha\leq0.1$. Moreover, the number density of created particles corresponding to point $C$ is even smaller than that without frequency modulation. This amazing result is caused by the same reason as the phenomenon occurring in Fig. \ref{fig:momentum}, where the highest momentum peak does not always correspond to pair creation by absorbing the photons with dominant frequency component. To explain this result, we give the number density of created pairs varying with the field frequency, see Fig. \ref{fig:range}. It can be seen that the pair number density is very sensitive to the field frequency. For example, the number density of created pairs by absorbing four photons with $\omega=0.512m$ is at least an order of magnitude greater than the one by absorbing four photons with $\omega=0.5m$. By analyzing the frequency spectra of the frequency modulated field at point $B$ (upper panel) and point $C$ (lower panel) shown in Fig. \ref{fig:point}, we find that the dominant frequency components at point $B$ and $C$ are $0.512m$ and $0.578m$, which correspond to a peak and a valley of the particle number density shown in Fig. \ref{fig:range}, respectively. Therefore, the pair number density at point $B$ is larger than the one without frequency modulation ($\omega=0.5m$), while the number density at point $C$ is smaller than the latter one. A similar result can also be found by comparing the values of momentum peaks in Figs. \ref{fig:momentum} (d), (a) and (f). Moreover, we know that the momentum peak $p_{18}$ in Fig. \ref{fig:momentum} (f) is formed by absorbing three photons with $\omega=0.578$ and one photon with $\omega=0.52$, and the number density for $\omega=0.52$ is much larger than that for $\omega=0.578$, so the value of momentum peak $p_{18}$ is higher than that of peak $p_{19}$, though the peak $p_{19}$ is formed by absorbing four photons with dominant frequency component. This analysis can also be used to understand momentum spectra in Figs. \ref{fig:momentum} (b) and (c), where highest peaks are not formed by absorbing four photons with dominant frequency component. By the way, when the modulation frequency at point $C$ changes from $0.01m$ to $0.009m$, one can see that the number density is also increased, see Fig. \ref{fig:number} (a), because the dominant frequency moves from $0.578m$ to $0.585m$ see Fig. \ref{fig:frequency} (b).

In the range of $0.32\leq\alpha\leq1.0$, the minimum and maximum values of the number density of created pairs are $E: 9.89\times 10^{-9}$ and $F: 2.03\times 10^{-5}$, respectively. It can be seen that the number density at point $F$ is $200$ times greater than that without frequency modulation ($N=1.04\times 10^{-7}$). It is noticeable that the value of $\alpha$ is $0.38$ at point $F$, which the effective frequency belongs to a reasonable area. Since the maximum effective modulation frequency should not be larger than the original laser frequency, the modulation parameters in the range of $\alpha\geq1.0$ have no actual meanings and are not discussed any more, although the number density of created particles is enhanced significantly in this region. Similar to Fig. \ref{fig:number} (a), the maximum value of the number density of created particles in Fig. \ref{fig:number} (b) is $L: 3.50\times 10^{-4}$, which is also $200$ times greater than the number density without frequency modulation ($N=1.74\times 10^{-6}$). In this case, the value of $\alpha$ is $0.576$ at point $L$ and the effective frequency is still in a relatively reasonable area.

Based on the above discussion, we find that the number density of created pairs is not always enhanced by a frequency modulated field, but for certain modulation parameters it indeed can be increased significantly. Therefore, according to our theoretical analysis the particle number density may be enhanced by over two orders of magnitude by conducting the most basic and precise modulation of a laser beam experimentally. It should be noted that the reason why the enhancement effect of pair production for the frequency modulated field here is much weaker than that for a linear chirped field is because the largest effective frequency of the frequency modulated field is restricted to less than two times the original frequency.

\section{Conclusion and discussion }
\label{sec:four}

In summary, we investigate the momentum spectrum and number density of created electron-positron pairs in frequency modulated electric field. For the momentum spectrum, it is found that there are obvious interference patterns on it. This interference effect can be well explained by analyzing the frequency spectra of the frequency modulated field. It finds that different momentum peaks correspond to the pair creation process by absorbing different frequency photons, because the frequency of photons for a frequency modulated field is not single. Moreover, we also provide another point of view to understand this interference effect by analyzing turning point structures. For the number density of created pairs, it is found that the number density of created pairs can be enhanced or weakened by the frequency modulated field, but for certain modulation parameters it can be improved by over two orders of magnitude. This result may provide a way to increase the number of created electron-positron pairs experimentally.


It is well known that this type of modulated laser field we considered is usually used to transmit information in communication system. In this paper, we find that the information carried by this field can also leave an imprint on the momentum spectrum of created pairs. Therefore, to some extent, we may obtain the information by analyzing the momentum spectrum. Furthermore, based on the recent work \cite{Su2019}, where the electric field with temporally modulated amplitude can be manipulated in a controlled way to transmit information from a sender to a receiver by a new transport medium for information ``Dirac vacuum", so it will be an interesting topic to check if the information carried by the frequency modulated electric field can be transmitted by the Dirac vacuum as well.

\acknowledgments
This work was supported by the National Natural Science Foundation of China under Grant No. 11705278 and No. 11875007, by the National Basic Research Program of China under Grant No. 2013CBA01504, by the National Key R\&D Program of China under Grant No. 2018YFA0404802, and partially by the Fundamental Research Funds for the Central Universities.


\begin{thebibliography}{99}

 \bibitem{Dirac1928}
 P. A. M. Dirac, Proc. Roy. Soc. Lond. A \textbf{117}, 610 (1928).

 \bibitem{Sauter1931}
 F. Sauter, Z. Phys. \textbf{69}, 742 (1931).

 \bibitem{Schwinger1951}
 J. S. Schwinger, Phys. Rev. \textbf{82}, 664 (1951).

 \bibitem{Piazza2012}
 For a comprehensive review, see A. Di Piazza, C. M\"{u}ller, K. Z. Hatsagortsyan, and C. H. Keitel, Rev. Mod. Phys. \textbf{84}, 1177 (2012).

 \bibitem{Xie2017}
 B. S. Xie, Z. L. Li, and S. Tang, Matter and Radiation at Extremes \textbf{2}, 225 (2017).

 \bibitem{Dunne2005}
 G. V. Dunne, and C. Schubert, Phys. Rev. D \textbf{72}, 105004 (2005).

 \bibitem{Dunne2006}
 G. V. Dunne, Q. H. Wang, H. Gies, and C. Schubert, Phys. Rev. D \textbf{73}, 065028 (2006).

 \bibitem{Lv2013}
 Q. Z. Lv, A. C. Su, M. Jing, Y. J. Li, R. Grobe and Q. Su, Phys. Rev. A \textbf{87}, 023416 (2013).

 \bibitem{Jing2013}
 M. Jiang, Q. Z. Lv, Z. M. Sheng, R. Grobe, and Q. Su, Phys. Rev. A \textbf{87}, 042503 (2013).

 \bibitem{Su2012}
 Q. Su, W. Su, Z. Q. Lv, M. Jing, X. Lu, Z. M. Sheng and R.Grobe, Phys. Rev. Lett. \textbf{109}, 253202 (2012).

 \bibitem{Gong2018}
 C. Gong, Z. L. Li, and Y. J. Li, Phys. Rev. A \textbf{98}, 043424 (2018).

 \bibitem{Kim2002}
 S. P. Kim, and D. N. Page, Phys. Rev. D \textbf{65}, 105002 (2002).

 \bibitem{Kim2007}
 S. P. Kim, and D. N. Page, Phys. Rev. D \textbf{75}, 045013 (2007).

 \bibitem{Alkofer2001}
 R. Alkofer, M. B. Hecht, C. D. Roberts, S. M. Schmidt, and D. V. Vinnik, Phys. Rev. Lett. \textbf{87}, 193902 (2001).

 \bibitem{Akkermans2012}
 E. Akkermans and G. V. Dunne, Phys. Rev. Lett. \textbf{108}, 030401 (2012).

 \bibitem{Hebenstreit2010}
 F. Hebenstreit, R. Alkofer, and H. Gies, Phys. Rev. D \textbf{82}, 105026 (2010).

 \bibitem{Li2014}
 Z. L. Li, D. Lu, B. S. Xie, L. B. Fu, J. Liu, and B. F. Shen, Phys. Rev. D \textbf{89}, 093011 (2014).

 \bibitem{Li2015}
 Z. L. Li, D. Lu, B. S. Xie, B. F. Shen, L. B. Fu, and J. Liu, Europhys. Lett. \textbf{110}, 51001 (2015).

 \bibitem{Li2017}
 Z. L. Li, Y. J. Li, and B. S. Xie, Phys. Rev. D \textbf{96}, 076010 (2017).

 \bibitem{Mocken2010}
 G. R. Mocken, M. Ruf, C. M\"{u}ller, and C. H. Keitel, Phys. Rev. A \textbf{81}, 022122 (2010).

 \bibitem{Akal2014}
 I. Akal, S. Villalba-Ch\'{a}vez, and C. M\"{u}ller, Phys. Rev. D \textbf{90}, 113004 (2014).

 \bibitem{Schutzhold2008}
 R. Schutzhold, H. Gies, and G. V. Dunne, Phys. Rev. Lett. \textbf{101}, 130404 (2008).

 \bibitem{Nuriman2012}
 A. Nuriman, B. S. Xie, Z. L. Li, and D. Sayipjamal, Phys. Lett. B \textbf{717}, 465 (2012).

 \bibitem{Dumlu2010}
 C. K. Dumlu, Phys. Rev. D \textbf{82}, 045007 (2010).

 \bibitem{Min2013}
 M. Jiang, B. S. Xie, H. B. Sang, and Z. L. Li, Chin. Phys. B \textbf{22}, 100307 (2013).

 \bibitem{Abdukerim2017}
 N. Abdukerim, Z. L. Li, and B. S. Xie, Chin. Phys. B \textbf{26}, 020301 (2017).

 \bibitem{Olugh2019}
 O. Olugh, Z. L. Li, B. S. Xie, and R. Alkofer, Phys. Rev. D \textbf{99}, 036003 (2019).

 \bibitem{Su2019}
 Q. Su and R. Grobe, Phys. Rev. Lett. \textbf{122}, 023603 (2019).

 \bibitem{Brezin1970}
 E. Brezin, and C. Itzykson, Phys. Rev. D \textbf{2}, 1191 (1970).

 \bibitem{Christian2014}
 C. Kohlf\"{u}rst, H. Gies, and R. Alkofer, Phys. Rev. Lett. \textbf{112}, 050402 (2014).

 \bibitem{Cesim2010}
 C. K. Dumlu, and G. V. Dunne, Phys. Rev. Lett. \textbf{104}, 250402 (2010).

 \bibitem{Cesim2011}
 C. K. Dumlu, and G. V. Dunne, Phys. Rev. D \textbf{83}, 065028 (2011).





\end{thebibliography}
\end{document}